\pdfoutput=1
\documentclass[10pt, aps,prd,amsmath,floats,floatfix, twocolumn,
superscriptaddress,nofootinbib,showpacs,longbibliography,a4paper]{revtex4-1}

\usepackage{amsmath,amssymb}
\usepackage{bm} 

\usepackage{soul}

\usepackage[T1]{fontenc}
\usepackage[utf8]{inputenc}
\DeclareUnicodeCharacter{2009}{\nobreakspace}

\usepackage{times}  

\usepackage[normalem]{ulem}
\usepackage{verbatim}
\usepackage[dvipsnames]{xcolor}
\definecolor{linkcolor}{rgb}{0.0,0.3,0.5}

\usepackage[
    hypertexnames=false,
    unicode,
    colorlinks=true,
    linkcolor=linkcolor,
    citecolor=linkcolor,
    filecolor=linkcolor,
    urlcolor=linkcolor,
    pdfusetitle
]{hyperref}

\usepackage{orcidlink}
\usepackage[all]{hypcap}
\usepackage{graphicx}
\usepackage{xspace}
\usepackage{microtype}
\usepackage[english]{babel}
\usepackage{mathrsfs}

\graphicspath{{images/}}

\newcommand{\Ft}{\widetilde{\mathcal{F}}(t)}
\newcommand{\Fw}{F(\mathit{w})}
\newcommand{\Ml}{M_{\ell}}
\newcommand{\Dl}{D_{\ell}}
\newcommand{\Ds}{D_{s}}
\newcommand{\Dls}{D_{\ell s}}
\newcommand{\los}{line of sight}

\begin{document}

\title{Probing the spin of compact objects with gravitational microlensing of gravitational waves}

\newcommand{\ICTS}{\affiliation{International Centre for Theoretical Sciences,
    Tata Institute of Fundamental Research, Bangalore 560089, India}}
\newcommand{\IACS}{\affiliation{Indian Association for the Cultivation of
    Science, 2A \& 2B Raja S C Mullick Road, Kolkata 700032, India}}
\newcommand{\IUCAA}{\affiliation{The Inter-University Centre for Astronomy and
    Astrophysics, Post Bag 4, Ganeshkhind, Pune 411007, India}}

\author{Gopalkrishna Prabhu\,\orcidlink{0009-0001-2695-3622}}
\email{gopal.prabhu@iucaa.in}
\IUCAA
\author{Uddeepta Deka\,\orcidlink{0000-0002-5942-4487}}
\email{uddeepta.deka@icts.res.in}
\ICTS
\author{Sumanta Chakraborty\,\orcidlink{0000-0003-3343-3227}}
\email{tpsc@iacs.res.in}
\IACS
\author{Shasvath J. Kapadia\,\orcidlink{0000-0001-5318-1253}}
\email{shasvath.kapadia@iucaa.in}
\IUCAA

\hypersetup{pdfauthor={Prabhu et al.}}

\date{\today}

\begin{abstract}
 Propagating gravitational waves (GWs) can encounter a massive object (lens) whose gravitational radius is comparable to the wavelength of the GWs (wave-optics regime). The resulting `microlensed' signal contains imprints about the properties of the lens. In this work, we compute the GW waveforms microlensed by a rotating compact object in weak-field gravity. Using these waveforms, for the first time, we assess how well the parameters of the rotating lens can be inferred from GW lensing observations. We find that if we allow for naked singular solutions within general relativity or beyond, which in principle can have spins that are not bounded to be extremal, our method can be used to extract the rotating lens parameters using observations of microlensed GWs with future detectors. As a result, we find that the lens parameters for such lenses are well recovered within $90\%$ confidence for signal-to-noise ratio (SNR) 50 and especially well for SNR$=100$ with Einstein Telescope. 
\end{abstract} 

\maketitle

\section{Introduction}\label{sec:intro}
Gravitational lensing of gravitational waves (GWs) occurs when GWs encounter large agglomerations of matter, such as massive black holes (BHs), galaxies, or clusters. While the LIGO-Virgo-KAGRA (LVK, \cite{aligo, avirgo, KAGRA}) network has completed multiple observing runs, and observed over two hundred compact binary coalescence (CBC) events via GWs \citep{LIGOScientific:2025slb, GWTC1, GWTC2, GWTC2.1, GWTC3}, none of these have been reported to exhibit confident signatures of lensing \cite{LIGOScientific:2023bwz}. Nevertheless, lensing is a highly anticipated observation, expected to be discovered for the first time in upcoming observing runs. The possible science that can be extracted from such observations is rich, shedding light on open questions in astrophysics, cosmology and fundamental physics. 

For example, GW lensing in the geometric optics regime -- which results in the production of multiple temporally resolved images -- can drastically enhance GW early warning \cite{Magare:2023hgs, Magare:2025ulm}, constrain cosmological parameters and properties of dark matter \cite{Jana:2022shb, Jana:2024dhc, Jana:2024uta, Barsode:2024wda}, probe the motion of Galactic neutron stars \cite{Basak:2022fig}, as well as the existence of polarization modes beyond those expected from General Relativity (GR) \cite{Goyal:2020bkm}, and provide precision measurements of the speed of GWs relative to the speed of light \cite{Collett:2016dey}. On the other hand, observation of lensing in the wave-optics regime, where a single image exhibiting interference, diffraction and beating patterns \cite{Takahashi2003}, can shed light on the properties of the lens (such as probing the existence of charge \cite{Deka:2024ecp}), constrain the fraction of dark matter as massive compact halo objects (MACHOs) \cite{Basak:2021ten}, and potentially shed light on the population of isolated black holes within galactic scale lenses \cite{Mishra:2021xzz}. In this work, we demonstrate that wave-optics modulations of GW signals emanated by stellar-mass CBCs can also, in principle, enable measurements of the spin of the lens.

We expect most astrophysical objects in the universe to have non-zero spin-angular momentum. The spin of compact objects is shaped by a rich set of physical processes which encodes the object's formation and evolutionary history. Gas clouds, protostellar cores, protoplanetary disks, and galaxies rarely begin perfectly motionless or perfectly symmetric. Even a very small initial rotation or random turbulent motion gives the gravitationally collapsing object some angular momentum. As the object contracts, conservation of angular momentum would increase the spin.  Conservation of angular momentum also implies that modest rotation in a progenitor star can produce substantial spin-up during core collapse \cite{Heger2005,Fuller2019}, leading to rapidly spinning compact objects. Moreover, remnants of CBCs are also expected to be rapidly spinning, due to the transfer of the binary's orbital angular momentum to the remnant. This is especially true of hierarchical merger remnants, which are expected to have large spins due to repeated transfer of orbital to spin angular momentum \citep{Gerosa2021}.

Over the last two decades, electromagnetic (EM) observations have developed several methods to infer the angular momentum of compact objects~\cite{McClintock:2013vwa, Fabian2000, Reynolds:2013qqa, vanderKlis1997, Akiyama2019, Eigenbrod2008, Pooley2007}. These EM approaches show that spin plays a first-order role in the dynamics, energetics and observed appearance of accreting systems, and that inference is often limited by astrophysical systematics, which can produce degeneracies with spin. As a result, EM constraints on spin are powerful but model dependent and, in many cases, systematically uncertain \cite{McClintock:2013vwa, Fabian2000, Banerjee:2021aln, Banerjee:2019nnj, Banerjee:2019sae, Mishra:2019trb}.

GW lensing could provide a complementary probe of the rotation of compact objects. When a lens possesses angular momentum, its gravitomagnetic field imprints a characteristic set of signatures on the propagating GWs. The gravitomagnetic field modifies the light-deflection angle and induces spin-dependent time delays, altering relative arrival times between images \cite{Kopeikin2002,Sereno2005, Asada:2000vn, Baraldo:1999ny}. In the wave-optics regime, these spin-induced time delays produce frequency-dependent fringe patterns whose phases and amplitudes are sensitive to the sign, magnitude and orientation of the lens spin. In the weak-field, slow-rotation regime, gravitomagnetic (spin) corrections enter the lensing potential as different powers of dimensionless spin. These corrections generate asymmetric deflections that depend on the path of the GW with respect to the spin angular momentum (prograde/retrograde), leading to shifts in image positions and relative time delays \cite{Asada:2000vn, Baraldo:1999ny,Sereno2003}.

In this work, we derive the wave-optics-driven modulations to the CBC's GW waveform due to a spinning lens. We assess, for the first time, the ability of future ground-based detectors to constrain the spin-parameter of the lens. The relevant parameter that induces spin-driven wave-optics phenomena is proportional to the Kerr-parameter, scaled by the typical size of the lens, i.e., its Einstein radius. This has two important implications. The first, from an analysis perspective it ensures that for a wide range of spin parameters, the slow-rotation limit (rotation parameter much smaller than the Einstein radius) can safely be invoked. But perhaps, most importantly, this scaling of the rotation parameter with the Einstein radius inhibits any meaningful probe of the spins of astrophysical BHs. Nevertheless, in general relativity (GR) as well as in a number of beyond GR theories, such as those that allow naked singularities, the spins are orders of magnitude larger than the Kerr limit and hence our method applies. We find that our method is able to strongly constrain such spins with next generation detectors such as Einstein Telescope (ET) \cite{Punturo:2010zz}. 

The rest of the paper is organised as follows. Section \ref{sec:methods} provides an overview of GW lensing, derives the bending angle, time-delay function and the lensing potential of a spinning lens in the weak field, slow-rotation limit, and summarizes the framework to produce waveforms lensed by spinning compact objects, as well as constrain the spin from the lensing-driven modulations. Section \ref{sec:results} reports our results of parameter estimation for example spinning lens systems. The paper ends with a summary and discussion in Section \ref{sec:summary}. 

\section{Lensing magnification due to a spinning compact object}\label{sec:methods}

In this Section, we first introduce the lensing magnification function and the physical quantities that define it. We then derive the lensing potential due to a spinning compact object. Finally, we describe the numerical method used to compute the lensing magnification function.

\subsection{GW lensing in the wave and geometric optics regimes}\label{GO_WO_lensing}

Lensing of GWs depends on the ratio between the GW wavelength, $\lambda_\mathrm{GW}$, and the gravitational length scale of a lens of mass $M_\ell$. For $\lambda_\mathrm{GW}\ll GM_\ell/c^2$, the geometric–optics limit applies. When the source and lens are well aligned along the line of sight, multiple lensed images can form, producing repeated signals separated by delays ranging from minutes to months, depending on the lens. This \textit{strong lensing} regime mirrors EM lensing: the lensed signals retain nearly identical phase evolution but differ by relative magnifications and constant phase shifts.

When $\lambda_\mathrm{GW}\sim GM_\ell/c^2$, wave-optics effects dominate, producing a single, frequency-modulated diffracted signal; this regime is referred to as \textit{microlensing}. Such diffraction signatures are essentially unobservable in EM lensing owing to their small wavelengths.


In this work, we consider the wave-optics microlensing of GW signals from a slowly rotating compact object in the weak field regime. Since in typical cosmological lensing scenarios, the size of the lens is expected to be much smaller than the distances between the source and the lens, we can use the thin-lens approximation, where the lens is assumed to be a two-dimensional planar surface instead of a three-dimensional object, whose plane is perpendicular to the \los. We also ignore the variation of the GW polarisation vector since it is suppressed by the smallness of the potential~\cite{Schneider:1992bmb}. The lensed GW in the frequency domain, $h_\ell$, can be written as the product of the unlensed signal, $h$, and the magnification function, $F$, as:
\begin{align}
    h_{\ell}(f) = F(f)\times h(f)\,.
\end{align}
In the thin-lens approximation, the magnification function $F(f)$ is obtained by evaluating the Fresnel diffraction integral over the entire lens plane~\cite{Schneider:1992bmb}:
\begin{equation}\label{eq:Fresnel integral}
F(f)
= \frac{\Ds \xi_0^2 (1+z_\ell)}{c\Dl \Dls}\;\frac{f}{i}\;
\int d^2\vec{x}\; \exp\!\big[2\pi i f\, t_d(\vec{x},\vec{\mathrm{y}})\big]\,,
\end{equation}
where $\vec{x}$ is the position vector in the lens plane in units of $\xi_0$, and $\vec{\mathrm{y}}$ is the position vector indicating the location of the source (also referred to as impact parameter) in units of $\Dl/(\xi_0 \Ds)$. Here $\xi_0$ is an arbitrary length scale, which will be fixed later. Further, $\Dl$ is the angular diameter distance from the observer to the lens, $\Ds$ is the angular diameter distance from the observer to the source and $\Dls$ is the angular diameter distance between the source and the lens, all along the line of sight (see Fig.~\ref{fig:lensing system}). The $(1+z_\ell)$ factor accounts for the cosmological redshift of the lens, with $z_\ell$ being the redshift of the lens. The time delay function (or the scaled Fermat potential), is given by \cite{Takahashi2003}:
\begin{equation}\label{eq:td4}
    t_d(\vec{x},\vec{\mathrm{y}})
    = \frac{\Ds \xi_0^2 (1+z_\ell)}{c\Dl \Dls} \left[ \frac{\lvert\vec{x}-\vec{\mathrm{y}}\rvert^2}{2} - \psi(\vec{x}) \right]\,,
\end{equation}
where $\psi(\vec{x})$ is the dimensionless two-dimensional deflection potential obtained by projecting the three-dimensional Newtonian potential onto the lens plane~\cite{Schneider1992}. This time delay can be viewed as originating from two mechanisms. The $\lvert\vec{x}-\vec{\mathrm{y}}\rvert^2/2$ factor is due to the geometric delay, i.e, the extra geometric path that the lensed GW has to travel with respect to the undeflected ray, and, the $\psi(\vec{x})$ part accounts for the time dialation effect due to the gravitational potential of the lens. We set the arbitrary length scale $\xi_{0}$ to be equal to the Einstein radius:
\begin{equation}
    \xi_0^2 = 4 G M_\ell \frac{\Dl \Dls}{c^2\Ds}\,.
\end{equation}

With this choice, the time delay function takes the form: 
\begin{equation}\label{timedelay2dpot}
    t_d(\vec{x},\vec{\mathrm{y}}) = \frac{4 G M_\ell}{c^3} (1+z_\ell) \times 
    \left[ \frac{|\vec{x}-\vec{\mathrm{y}}|^2}{2} - \psi(\vec{x}) \right]\,.
\end{equation}
Defining the dimensionless frequency $\mathit{w}$ as:
\begin{subequations}\label{eq:dimensionless_frequency}
    \begin{align}
        w &\equiv \frac{\Ds}{c\Dl \Dls}\xi_0^2(1+z_\ell)2\pi f \\
        & = \frac{4 G M_\ell}{c^3}(1+z_\ell)2 \pi f ~,
    \end{align}
\end{subequations}
the magnification function in Eq.~\eqref{eq:Fresnel integral} can be written in a simpler form as:
\begin{equation}
  \label{eq:ampfac_dimensionless_fd}
  \Fw = \frac{\mathit{w}}{2\pi i}\int d^{2}\vec{x}\;\exp\left[i\mathit{w}\tau_{\rm d}(\vec{x},\vec{\mathrm{y}})\right],
\end{equation}
where $\tau_{\rm d}$ is the dimensionless time delay, given by:
\begin{equation}
  \label{eq:time_delay_dimensionless}
  \tau_{\rm d}(\vec{x},\vec{\mathrm{y}}) = \frac{|\vec{x}-\vec{\mathrm{y}}|^{2}}{2}-\psi(\vec{x}).
\end{equation}

The main roadblock in computing the modulation $F(\mathit{w})$ in Eq.~\eqref{eq:ampfac_dimensionless_fd} is the integral involving the time delay function, which has an analytically closed form solution only for a few spherically symmetric lenses, e.g., the non-rotating point mass lens \cite{Takahashi2003}. For a few other cases, such as that of a charged lens \cite{Deka:2024ecp}, the time delay function can be determined analytically, but the magnification function had to be obtained numerically following \cite{Deka:2025vzx} (see Appendix~\ref{sec:appex_A:PR} for details). Here we will show that the same is true for the case of a slowly rotating lens, for which also there is no exact closed-form solution for the magnification function, requiring numerical methods.

In the geometric-optics regime, the total field can be approximated by a superposition of contributions from discrete stationary points of the Fermat potential, i.e., only the stationary points of the Fermat potential contribute to Eq.~\eqref{eq:ampfac_dimensionless_fd}. Thus the lensed GWs can be treated as separate, temporally resolved, copies of the unlensed GWs. It turns out that the stationary points play a crucial role in determining the frequency modulated waveform, and these points can be obtained from the time delay function as:
\begin{equation}
{
    \vec{\nabla}_{\vec{x}}\,t_d(\vec{x},\vec{\mathrm{y}}) \Big|_{\vec{x}=\vec{x}_j} =0\,~.
}
\end{equation}
From Eq.~\eqref{timedelay2dpot}, it follows that in the geometrical optics limit the lens mapping equation becomes:
\begin{equation}\label{eq:lens_mapping}
{
    \vec{\mathrm{y}} = \vec{x} - \vec{\nabla}\psi(\vec{x})~,
} 
\end{equation}
whose solution for a fixed impact parameter $\mathbf{y}$ gives rise to a set of solutions $\mathbf{x}_j$, acting as an image.
In the present context we use the above equations to get approximate image locations, which is useful in the numerical computation of the magnification function.

\subsection{Derivation of lensing observables of a slowly rotating compact object}\label{sec:lensing-potential-derivation}

We derive the effective Newtonian potential, deflection angle, and lensing potential for a slowly rotating compact object, retaining up to linear order terms in the angular momentum and working in the weak-field limit, i.e, keeping terms upto $\mathcal{O}(G)$. 

For this purpose we write down the post-Minkowskian metric for a slowly rotating object, which can be expressed as~\cite{Misner:1973prb}:
\begin{align}\label{eq:slow-rotation-metric}
    ds^2 = & -\left(1 + \frac{2U}{c^2}\right) c^2dt^2 + \left(1 - \frac{2U}{c^2}\right)\delta_{ij} \,dx^i dx^j\nonumber \\
    & + \frac{2}{c^2}V_i dx^i dt,
\end{align}
where the scalar and vector gravitational potentials are given by:
\begin{equation}\label{eq:grav-potentials}
    U(r) = -\frac{G\Ml}{r}, \quad \vec{V}(\vec{r}) = -\frac{2G}{r^3}(\vec{J} \times \vec{r}).
\end{equation}
Here, $\Ml$ and $\vec{J}$ are the mass and angular momentum of the lens, respectively, and $\vec{r}$ is the spatial position vector with magnitude $r = |\vec{r}|$, which is centered at the object causing the lensing.

The deflection angle for GWs propagating from source to observer takes the following form in the weak gravity regime~\cite{Asada:2000vn}:
\begin{align}\label{eq:deflection-angle-definition}
    \hat{\vec{\alpha}}&=\int_{\rm source}^{\rm observer}\Big[\frac{2}{c^2}\left\{\vec{\nabla} - \hat{n}(\hat{n}\cdot\vec{\nabla})\right\}U 
    \nonumber
    \\
    &\qquad-\frac{1}{c^3}\hat{n} \times (\vec{\nabla} \times \vec{V})\Big] dl\,,
\end{align}
where $l$ parametrizes the path of the lensed GW from the source to the observer. To simplify the setting further, we implement the Born approximation, or equivalently, the small deflection angle approximation. In this scheme the same expression as Eq.~\eqref{eq:deflection-angle-definition} holds, while $l$ now parameterizes the unperturbed light path, such that $D_{\ell s}=\int dl$ and $\vec{r}(l) = \vec{\xi} + l\hat{n}$, where $\hat{n}$ is the direction of propagation of the GWs.

\begin{figure}[t!]
    \centering
    \includegraphics[width=1\linewidth]{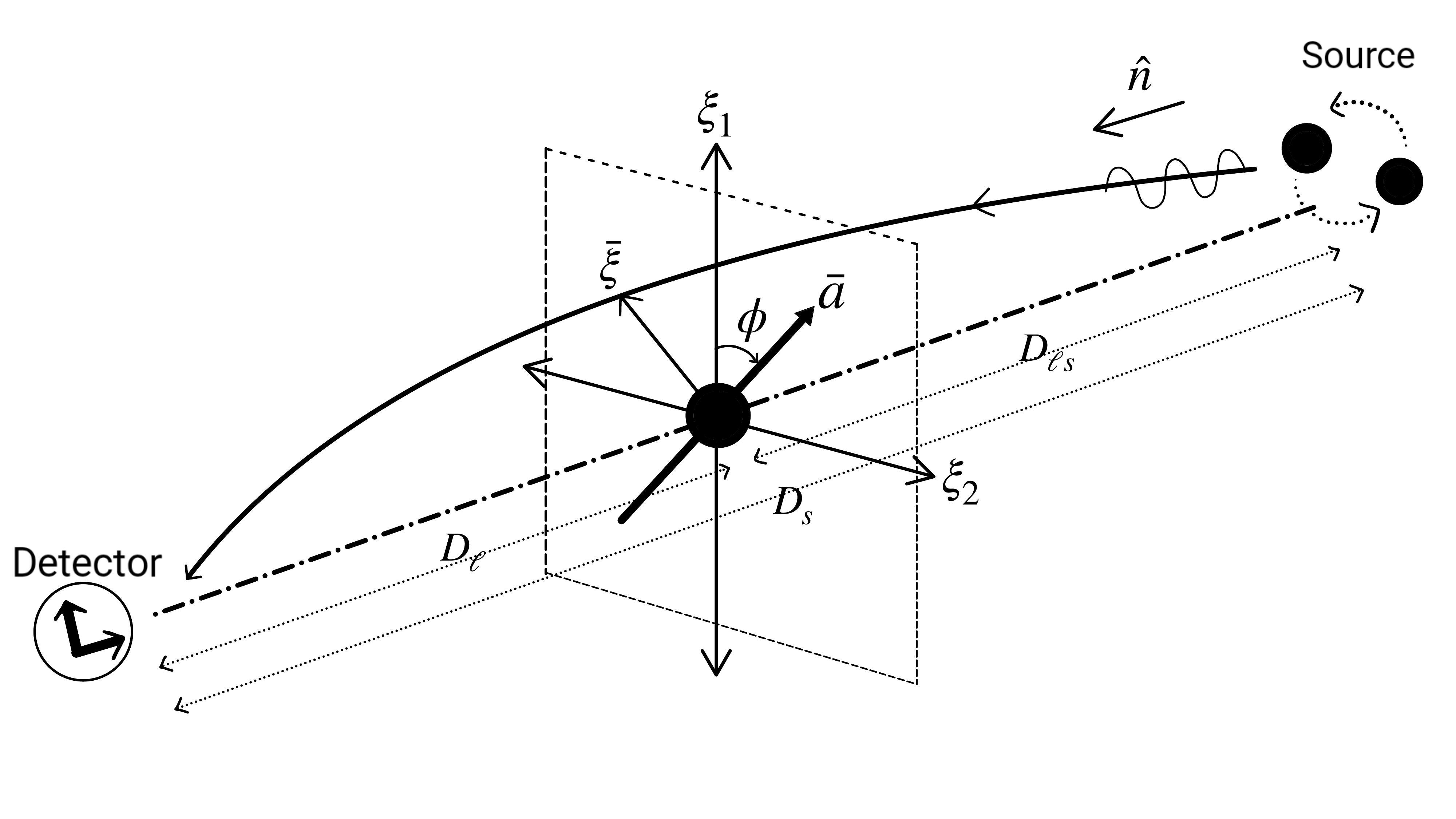}
    \caption{Schematic describing the coordinate system for lensing of a GW due to a compact object with spin $\bar{a}$ in the thin lens approximation. The GW propagating in the direction $\hat{n}$ strikes the lens plane ($\xi_1-\xi_2$) at $\bar{\xi}$, and travels to the detector after being lensed. The \los is shown as the dashed-dotted line perpendicular to the lens plane. The spin $\bar{a}$ is assumed to lie entirely in the lens plane, making an angle $\phi$ with the $\xi_1$ axis. $\Dl$, $\Ds$ and $\Dls$ are the angular diameter distances from the detector to the lens, from the detector to the source and from the lens to the source, respectively, along the \los.}
    \label{fig:lensing system}
\end{figure}

Simplifying Eq.~\eqref{eq:deflection-angle-definition} using three dimensional vector identities, the deflection angle of the GW becomes:
\begin{align}\label{eq:deflection-angle-born-approximation}
    \hat{\vec{\alpha}} &= \int_{\rm source}^{\rm observer} \frac{2}{c^2} \vec{\nabla}\left(U - \frac{1}{2c} \hat{n} \cdot \vec{V} \right) dl \nonumber \\
    &\quad - \int_{\rm source}^{\rm observer} (\hat{n} \cdot \vec{\nabla}) \left( \frac{2}{c^2} \hat{n} U - \frac{1}{c^3} \vec{V} \right) dl.
\end{align}
The second term is a total derivative, as $\hat{n}.\vec{\nabla}=(d/dl)$ and hence vanishes under the assumption of asymptotic flatness at the location of the source and the observer:
\begin{equation}
    \left[ \frac{2}{c^2} \hat{n} U - \frac{1}{c^3} \vec{V} \right]_{\rm source}^{\rm observer} = 0\,.
\end{equation}
Hence, the deflection angle reduces to:
\begin{equation}\label{eq:deflection-angle-integral-form}
    \hat{\vec{\alpha}} \approx \int_{-\infty}^{\infty} \frac{2}{c^2} \vec{\nabla} \widetilde{U} \, dl\,,
\end{equation}
with the effective potential:
\begin{equation}\label{eq:effective-Newtonian-potential}
    \widetilde{U} = U - \frac{1}{2c} \hat{n} \cdot \vec{V} = -\frac{G\Ml}{r} - \frac{G}{cr^3} \vec{r} \cdot (\vec{J} \times \hat{n})\,.
\end{equation}

Using cylindrical coordinates $\{\hat{\xi}, \hat{\phi}, \hat{n}\}$ (see Fig.~\ref{fig:lensing system}), the deflection angle becomes:
\begin{equation}\label{eq:deflection-angle}
    \hat{\alpha}(\vec{\xi}) = \frac{4G\Ml}{c^2\xi}\left[\left(1 + \frac{\vec{a}\cdot\hat{\phi}}{c\xi}\right)\hat{\xi} + \frac{\vec{a} \cdot \hat{\xi}}{c\xi} \hat{\phi} \right],
\end{equation}
where $\vec{a} = \vec{J}/\Ml$ is the specific angular momentum vector.

We define scaled coordinates $\vec{x} = \vec{\xi}/\xi_0$ and dimensionless spin $\hat{a} = \vec{a}/(c \xi_0)$. The scaled deflection angle in these quantities then reads:
\begin{subequations}\label{eq:scaled-deflection-angle}
\begin{align}
    \vec{\alpha}(\vec{x}) &\equiv \frac{\Dl\Dls}{\xi_0\Ds} \hat{\alpha}(\xi_0\vec{x}) \\
    &= \frac{1}{x} \left[\left(1 + \frac{\hat{a} \cdot \hat{\phi}}{x} \right)\hat{x} + \frac{\hat{a} \cdot \hat{x}}{x} \hat{\phi} \right]
\end{align}
\end{subequations}
where $x = |\vec{x}|$. It can be checked that $\vec{\nabla} \times \vec{\alpha} = 0$. This implies that the deflection angle is conservative, and we can define a lensing potential $\psi$ such that $\vec{\alpha} = \vec{\nabla} \psi$. Integrating, we obtain the lensing potential for the spinning lens as:
\begin{equation}\label{eq:lensing-potential}
    \psi(\vec{x}) = \ln x - \frac{1}{x^2} \vec{x} \cdot (\hat{a} \times \hat{n})
\end{equation}
where the second term encodes the spin-induced asymmetry.
The dimensionless time delay in Eq.~\eqref{eq:time_delay_dimensionless} then becomes

\begin{equation}
    \tau_\mathrm{d}(\vec{x}) = \frac{1}{2}|\vec{x} - \vec{\mathrm{y}}|^2 - \ln x + \frac{1}{x^2} \vec{x} \cdot (\hat{a} \times \hat{n}).
\end{equation}

The spin-dependent correction implies that prograde (retrograde)\footnote{Prograde (retrograde) paths refer to when the angular momentum of the GW is in the same (opposite) direction as the spin of the lens} paths are deflected less (more) as compared to the Schwarzschild deflection alone\footnote{This also means that prograde GWs experience shorter time delays than retrograde GWs}, which are in agreement with the results in \cite{iyer2009strongweakdeflectionlight, Asada:2000vn}. Note that this is contrary to what is implied if we use the time delay equation, Eq. (20) in \cite{Nakamura1998}, which implies that the time delay is more (less) for prograde (retrograde) paths as compared to the Schwarzschild deflection time delay.

\subsection{Caustics, critical curves and number of images}

\begin{figure*}[t!]
    \centering
    \includegraphics[width=\textwidth]{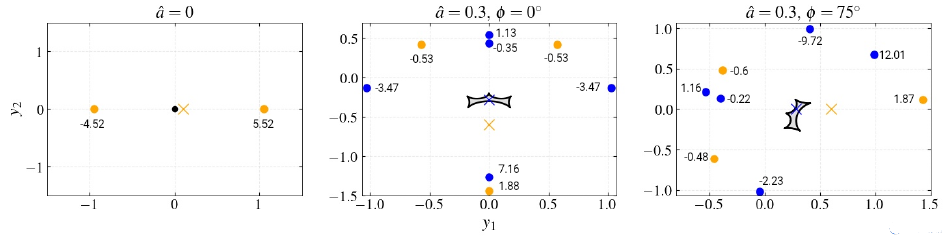}
    \caption{ Caustic curves in the source plane ($y_1-y_2$) for a non-rotating (left) versus a rotating lens (middle $\&$ right) with dimensionless spin parameter $\hat{a}=0.3$ for $\phi$ fixed at $\phi=0^\circ$ and $\phi=75^\circ$, respectively. The caustic for a non-rotating point mass lens is a single point at the origin (black dot), and we always have two images (orange dots) for the source (orange cross). In the case of a rotating lens, if the source is inside the grey shaded region (blue cross), lensing produces five images (blue dots), whereas if the source lies anywhere outside the shaded region (orange cross), there are three images (orange dots). Also shown are the GO image magnifications for individual images (black numbers). As the angle $\phi$ between the $\mathrm{y}_1$ axis and $\hat{a}$ changes, the caustic for the rotating lens rotates, and the corresponding image locations shift.}
    \label{fig:Caustics}
\end{figure*}

The number of classical lensed images in the GO limit is equal to the number of solutions of the lens equation (Eq. \eqref{eq:lens_mapping}). For a Schwarzschild lens, there are two classical lensed images, whereas, for a lens with spin (linear order), there can be up to 5 images depending on the source location~\cite{Bonga:2024orc}. A useful tool to understand the number of images as a function of the location of the source are caustics and critical curves. Caustics are curves in the source plane where the image magnification formally diverges in the GO limit. They also define boundaries in the source plane between regions of different number of lensed images \cite{Schneider:1992bmb}. The critical curve in the lens plane corresponding to it's caustic in the source plane can be found by using the lens mapping equation Eq.~\eqref{eq:lens_mapping}. In Fig.~\ref{fig:Caustics}, we show the caustic curves in the case of a rotating lens with dimensionless spin parameter $\hat{a}=0.3$.
Inside the grey shaded region (caustic), the number of images is five, and outside it, the number of images is three. We also show, for comparison, the caustic curve for a non-rotating point lens for which the caustic is a point at the source plane origin. In each case, we also show examples of the source locations and their corresponding images. It is interesting to note that observation of five lensed images can be a tell-tale sign of lensing via a rotating BH with such  spin-parameters.\footnote{There are no standard lens models which can produce five classical images, to the best of our knowledge.}

\subsection{Numerical calculation of the magnification function}
\begin{figure*}[ht]
    \centering
    \includegraphics[width=\textwidth]{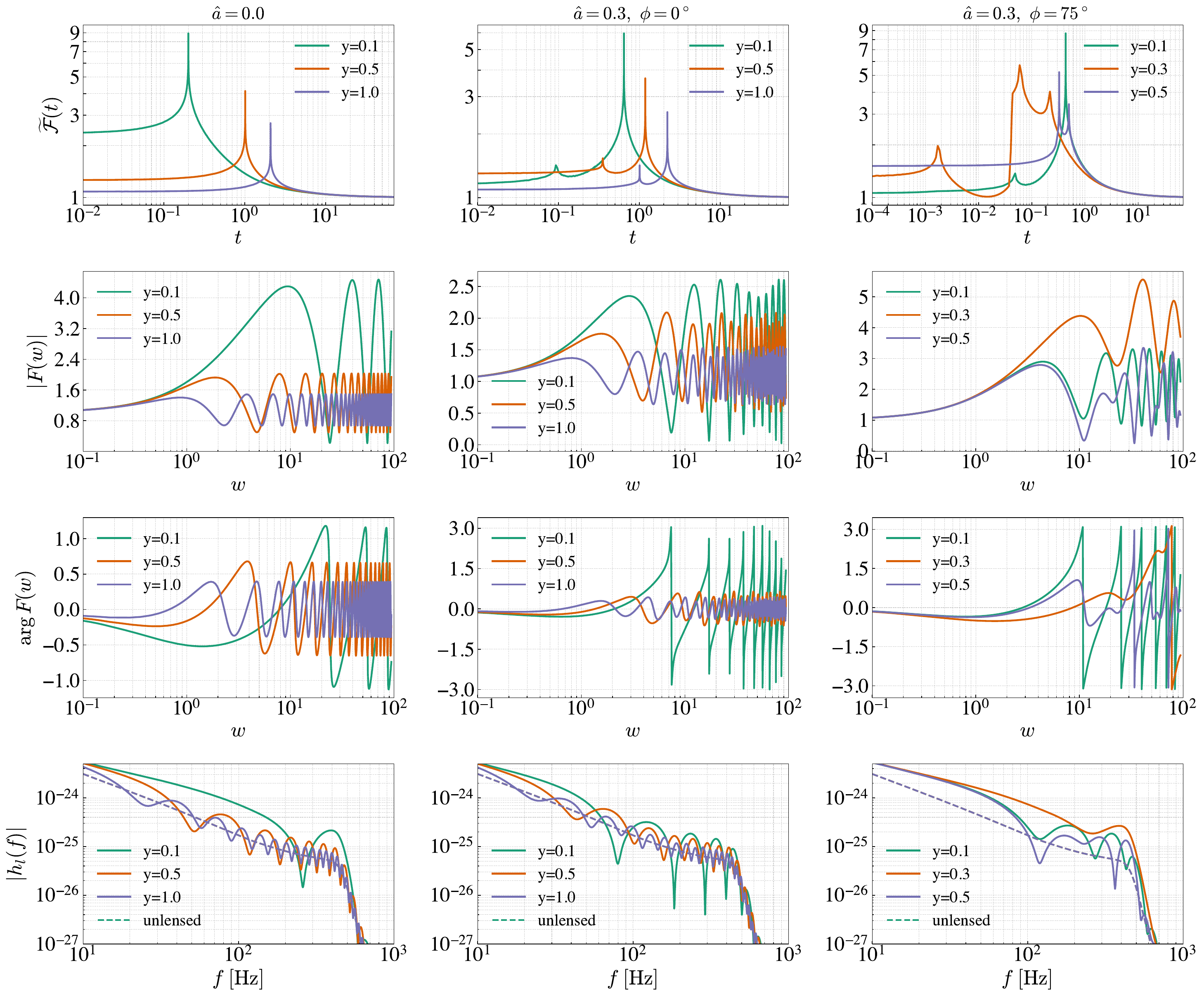}
    \caption{Top panel: Time-domain magnification function $\Ft$ as a function of $t$ (in units of $4M_\ell (1+z_\ell)$, with the global minima at $t=0$). The columns indicate various values of the spin parameter $\hat{a}=\{0, 0.3\}$ and the spin angle $\phi=\{0^\circ, 75^\circ \}$. Colours indicate the source position $\mathrm{y}=\{0.1, 0.5, 1.0\}$. Second panel: Amplitude of the magnification function, $|\Fw|$ as a function of the dimensionless frequency $w$. For the cases with non-zero spin, $|\Fw|$ is significantly modulated away from the non-rotating case. Third panel: Phase of the amplification function arg$\Fw$, as a function of  $w$. Bottom panel: Amplitude of the lensed GW strain as a function of frequency $f$ in Hz. In all cases, the lens mass is $M_\ell=300M_\odot$ located at redshift $z_\ell=0.5$. The unlensed waveform is also shown for reference, which is a quasi-circular waveform from a $20-20 M_\odot$ binary at redshift $z_s=2$.}
    \label{fig:waveforms_combined}
\end{figure*}

As mentioned earlier, the Fresnel integral Eq.~\eqref{eq:ampfac_dimensionless_fd} does not have an analytically closed solution except for a few simple lens models such as the point-mass lens model \cite{Schneider1992}. Since for a rotating lens, Eq.~\eqref{eq:ampfac_dimensionless_fd} does not have a closed-form solution, we use numerical methods for its calculation. Direct numerical calculation of Eq.~\eqref{eq:ampfac_dimensionless_fd} using standard numerical integration methods can be problematic because of the highly oscillatory integrand for large $\mathit{w}\tau_d$. However, there are several clever numerical methods that can be used to calculate the magnification function (see for example \cite{waveoptics_numerical_methods}). In this work, we calculate the magnification function following the method used in \cite{Deka:2024ecp, Deka:2025vzx}, with some modifications. In summary, we first calculate the magnification function in the time domain and then calculate the frequency domain magnification function using Fourier transforms. The time domain magnification function diverges logarithmically at the times corresponding to the saddle images. Hence, numerical computations require extra resolution around these points. To improve the numerical accuracy around such points, we use an analytic approximation of the saddle peak. For a detailed description of these techniques, please refer to Appendix \ref{sec:Numcal_amp_fac} and \cite{Deka:2024ecp, Deka:2025vzx}. The resulting magnification functions and waveforms using this numerical method are shown in Fig.~\ref{fig:waveforms_combined}.
In the top panel we show the time-domain magnification function $\Ft$ as a function of $t$ (in units of $4M_\ell (1+z_\ell)$, with the global minimum at $t=0$). The columns indicate various values of the spin parameter $\hat{a}=\{0, 0.3\}$ and the spin angle $\phi=\{0^\circ, 75^\circ \}$, for various values of source positions $\mathrm{y}$. The logarithmically diverging peaks correspond to saddle images. For the non-rotating point mass lens, we always get two images: a global minimum and a saddle image. For the rotating lens, the spherical symmetry of the non-rotating point-mass lens is broken and we can have upto five images, depending on whether the source lies inside or outside the caustic (see Fig. \ref{fig:Caustics})
The second and third panels show the amplitude and the phase of the magnification function, $|\Fw|$ and arg$\Fw$, respectively, as a function of the dimensionless frequency $w$. For the case with non-zero spin, both $|\Fw|$ and arg$\Fw$ are significantly modulated away from the non-rotating case. Also shown in the bottom panel is the GW amplitude of the lensed GW as a function of frequency $f$ in Hz. In all cases, the lens mass is $M_\ell=300M_\odot$ located at redshift $z_\ell=0.5$. The unlensed waveform is also shown for reference, which is a quasi-circular waveform from a $20-20 M_\odot$ binary at redshift $z=2$.

\section{Prospective constraints on the spins of compact objects from lensing observations}\label{sec:results}

In this Section, we present the forecasted constraints on the spin of compact objects from future observations of microlensed GW signals and the formalism to compute them. As seen in Fig.~\ref{fig:waveforms_combined}, lensed waveforms with non-zero spins exhibit morphologies that are clearly distinct from the zero-spin, or PML, case. We make use of a Bayesian framework to quantify these differences, perform parameter estimation (PE), and assess the measurability of spin-induced lensing effects.

The microlensed GW signal is described by a set of source parameters $\vec{\theta}_\mathrm{src}$ and lens parameters $\vec{\theta}_\ell$. Given observed data $d$, the joint posterior distribution of these parameters follows from Bayes' theorem:
\begin{equation}
    p(\vec{\theta}_{\rm src}, \vec{\theta}_\ell \,|\, d) =
    \frac{p(\vec{\theta}_{\rm src}, \vec{\theta}_\ell)\,
      p(d \,|\, \vec{\theta}_{\rm src}, \vec{\theta}_\ell, \mathcal{H}_{\mathrm{ML}})}
     {p(d \,|\, \mathcal{H}_{\mathrm{ML}})} ,
    \label{Eq:joint posterior}
\end{equation}
where $p(\vec{\theta}_{\rm src}, \vec{\theta}_\ell)$ is the prior on the source and lens parameters. The term $p(d \,|\, \vec{\theta}_{\rm src}, \vec{\theta}_\ell, \mathcal{H}_{\mathrm{ML}})$ is the likelihood of the data under the microlensing hypothesis $\mathcal{H}_{\mathrm{ML}}$, i.e., assuming $d$ contains a microlensed GW signal. The denominator $p(d \,|\, \mathcal{H}_{\mathrm{ML}})$ is the likelihood marginalised over all the parameters, and it serves as the normalisation constant for the posterior. 

For simplicity, we assume that the source parameters $\vec{\theta}_\mathrm{src}$ are largely uncorrelated with the lens parameters $\vec{\theta}_\ell$. This is a reasonable approximation, although recent studies have shown that microlensing-induced modulations can mimic features produced by spin-induced precession of the binary source~\cite{Mishra:2023ddt}. We also neglect the possibility that alternative lens models -- such as a charged lens~\cite{Deka:2024ecp} or a non-spinning point-mass lens embedded in a macrolens potential~\cite{Diego:2019lcd, Meena:2019ate} -- can produce waveform modulations similar to those generated by a spinning lens. Further, we fix the projected spin-angle $\phi$ to its true value to limit the dimensionality of the lens parameter space and focus on the dominant spin lens effects. For simplicity, we also assume that the source parameters are fixed to their true values.

Under these assumptions, the posterior on the lens parameters in Eq.~\eqref{Eq:joint posterior} reduces to 
\begin{equation}
p(\vec{\theta}_\ell \,|\, d) \propto
p(\vec{\theta}_\ell)\;
p(d \,|\, \vec{\theta}_\ell, \mathcal{H}_{\mathrm{ML}})~,
\label{Eq:joint posterior src fixed}
\end{equation}
where the relevant lens parameters are $\vec{\theta}_\ell = \{\hat{a}, M_\ell, y \}$. 

A full Bayesian PE would require evaluating the likelihood at multiple points in the parameter space within a reasonable time. However, generating fast microlensed waveforms for our model is computationally prohibitive. Therefore, we restrict ourselves to the high signal-to-noise ratio (SNR) limit, in which case the expectation value of the likelihood can be approximated as (see Appendix~\ref{sec:lkhd_high-SNR_justification} for details): 
\begin{equation}\label{eq:lkhd_high-SNR}
    \mathcal{L}(\vec{\theta}_\ell) 
    \equiv \langle p(d \,|\, \vec{\theta}_\ell, \mathcal{H}_{\mathrm{ML}}) \rangle \propto \exp \left[ -\rho^{2} \mathcal{M}(\vec{\theta}_\ell^{\,\mathrm{tr}},\vec{\theta}_\ell) \right] .
\end{equation}
Here, $\rho$ is the SNR of the signal in the data and $\mathcal{M}$ is the mismatch between two waveforms: the true waveform $h_\ell(f;\vec{\theta}_\ell^\mathrm{tr})$ with injected parameters $\vec{\theta}_\ell^\mathrm{tr}$, and a template waveform $h_\ell(f;\vec{\theta})$, with trial parameters $\vec{\theta}_\ell$~\footnote{Note that we set the source parameters of the template waveforms equal to those of the true waveform.}. The mismatch, which quantifies the dissimilarity between two waveforms, is defined as:
\begin{equation}\label{eq:match_definition}
    \mathcal{M}(\vec{\theta}_\ell^{\,\mathrm{tr}},\vec{\theta}_\ell)\equiv1-\max_{\varphi_c, \,t_c}\frac{\langle h_\ell(\vec{\theta}_\ell^\mathrm{tr}), h_\ell(\vec{\theta}_\ell) \rangle}{\sqrt{\langle h_\ell(\vec{\theta}_\ell^\mathrm{tr}), h_\ell(\vec{\theta}_\ell^\mathrm{tr})\rangle\langle h_\ell(\vec{\theta}_\ell), h_\ell(\vec{\theta}_\ell) \rangle}}~,
\end{equation}
where the maximisation is done over the coalescence phase $\varphi_c$ and the coalescence time $t_c$. The inner product $\langle\cdot\rangle$ between two waveforms $h(\vec{\theta}_1)$ and $h(\vec{\theta}_2)$ is defined as:
\begin{equation}\label{eq:inner_product}
    \langle h(\vec{\theta}_1), h(\vec{\theta}_2)\rangle\equiv 4 \,{\rm{Re}}\int_{f_{\rm{low}}}^{f_{\rm{upp}}}\frac{h^\ast(f;\vec{\theta}_1) \,h(f;\vec{\theta}_2)}{S_n(f)} df.
\end{equation}
Here $^\ast$ denotes complex conjugation. The detector bandwidth is set by the lower and upper frequency cutoffs $f_{\rm{low}}$ and $f_{\rm{upp}}$, respectively. $S_n(f)$ is the one-sided noise power spectral density (PSD). For this work, we make use of the ET-D detector sensitivity~\cite{Hild2011ETsensitivity}. This form of the likelihood in Eq.~\eqref{eq:lkhd_high-SNR} shows that, in the high-SNR limit, it is sharply peaked around the minimum mismatch between the template and the true signal, with the peak width determined by the SNR.

\begin{figure*}[ht]
  \centering
  \begin{tabular}{cc}
    \includegraphics[width=0.48\linewidth]{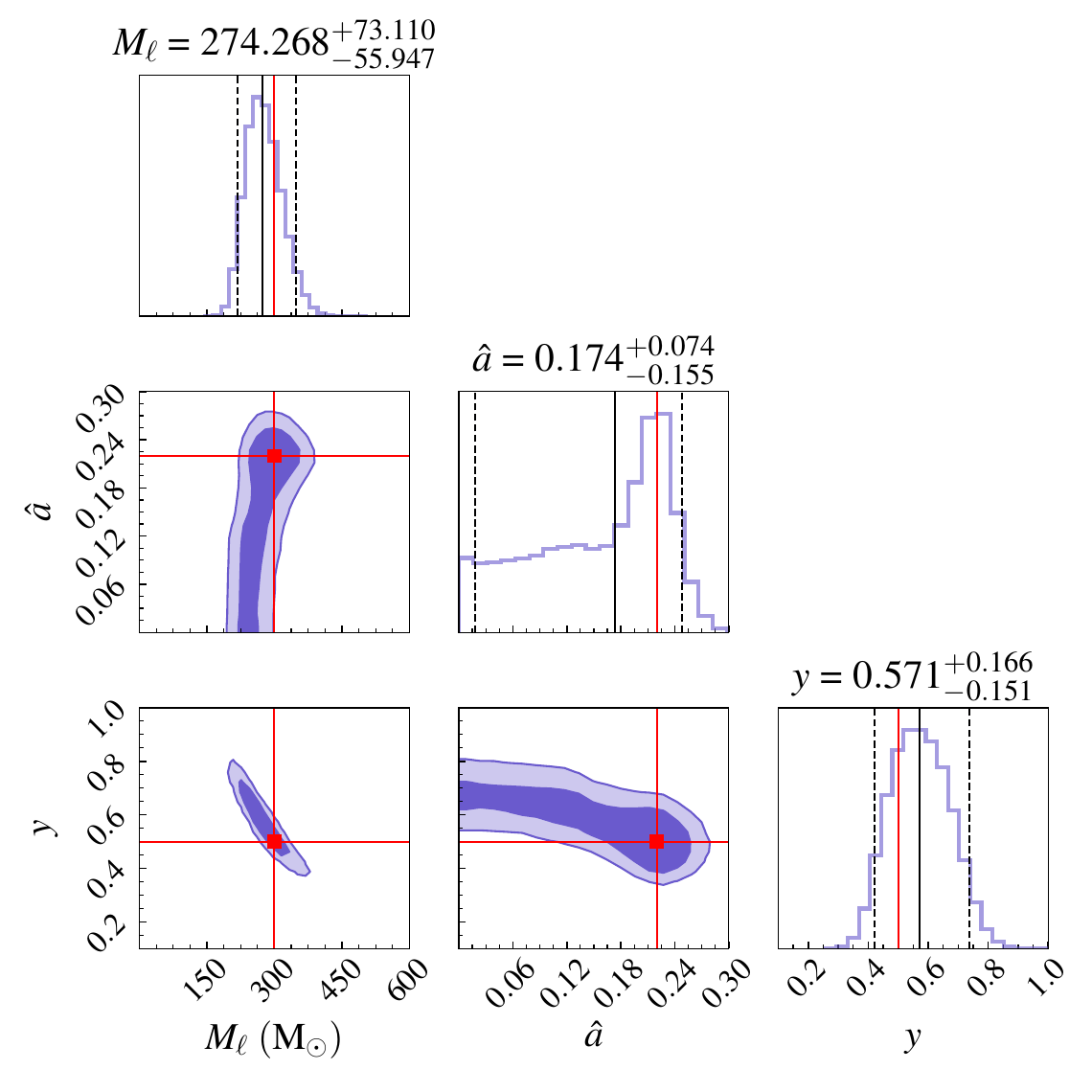} &
    \includegraphics[width=0.48\linewidth]{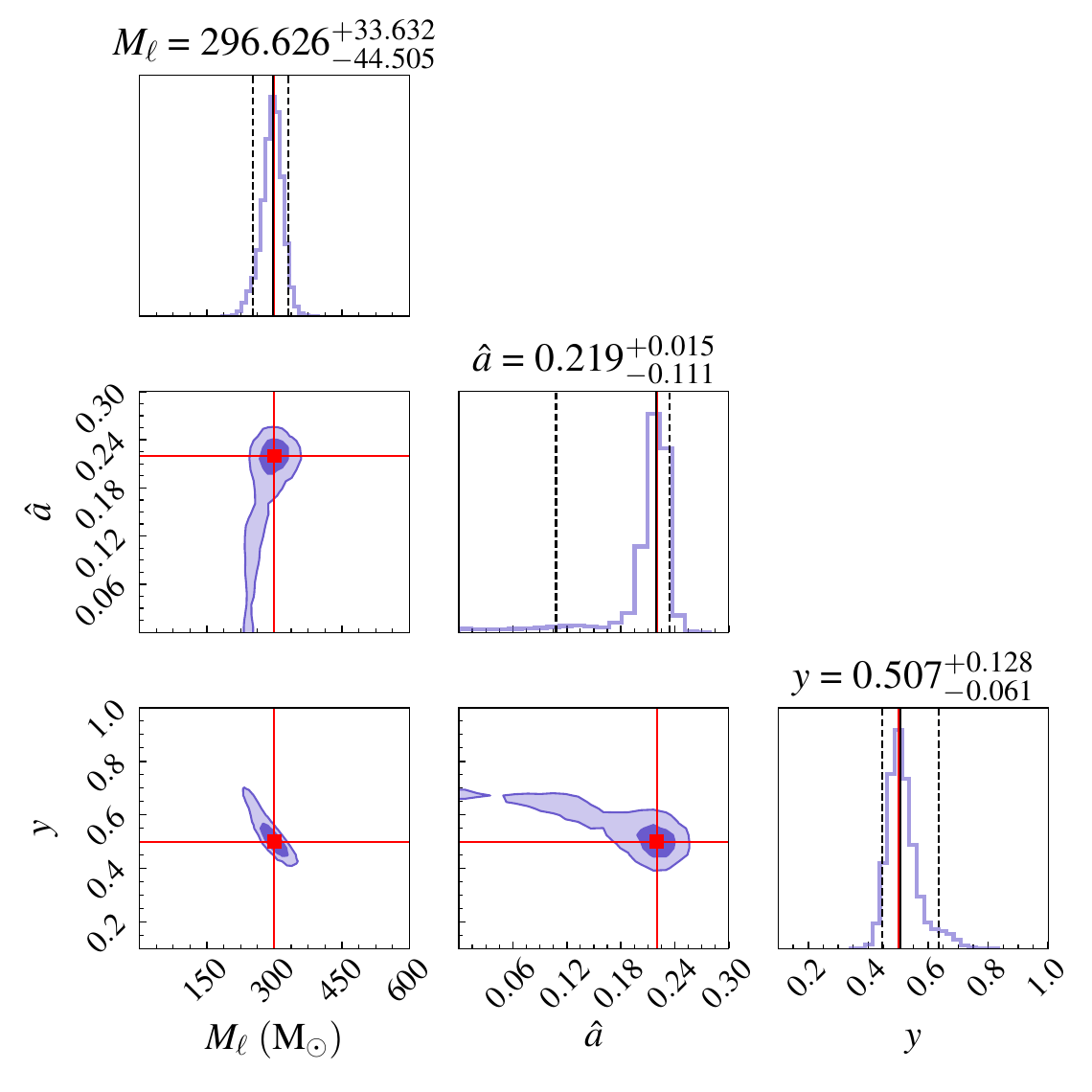} \\[0.5cm]
    \includegraphics[width=0.48\linewidth]{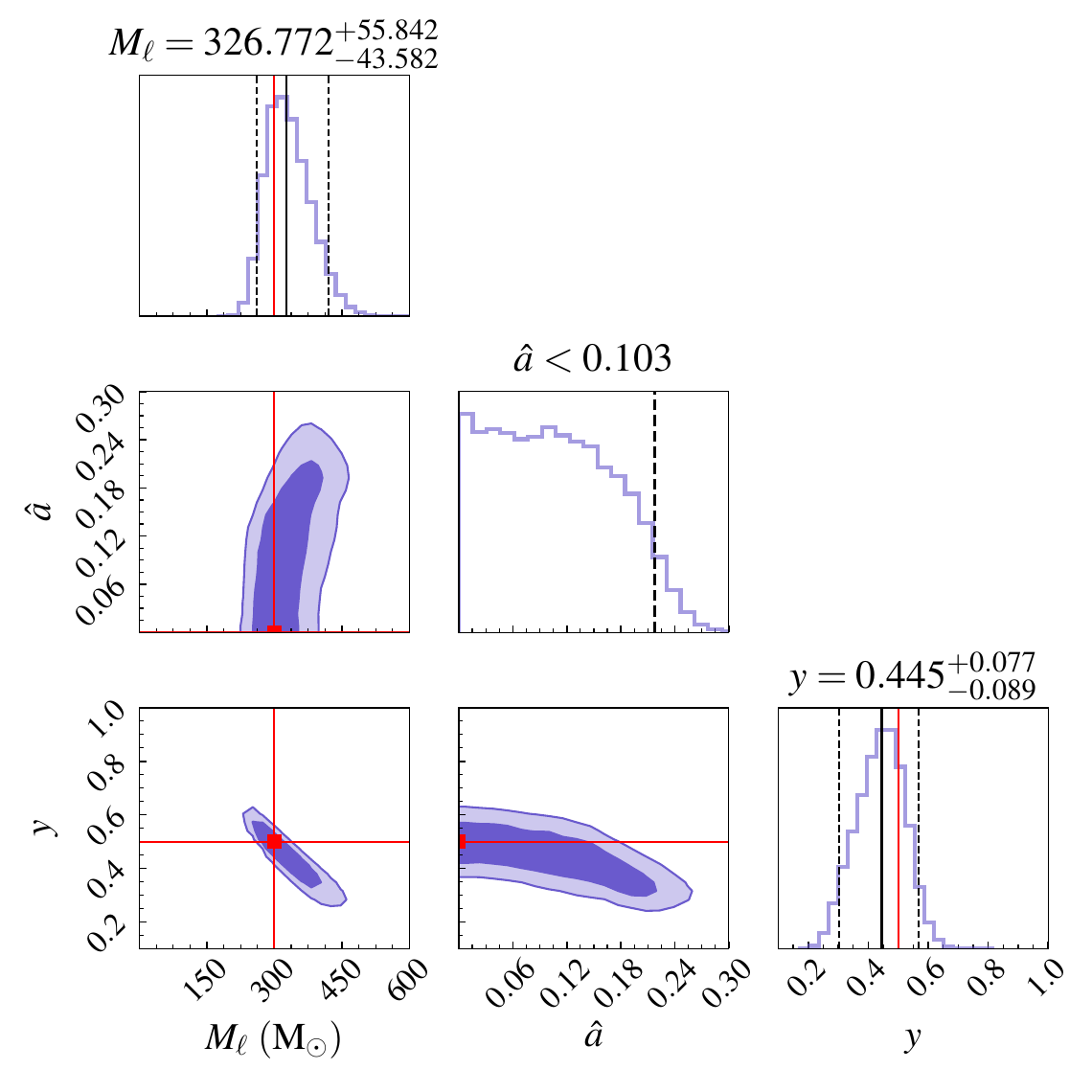} &
    \includegraphics[width=0.48\linewidth]{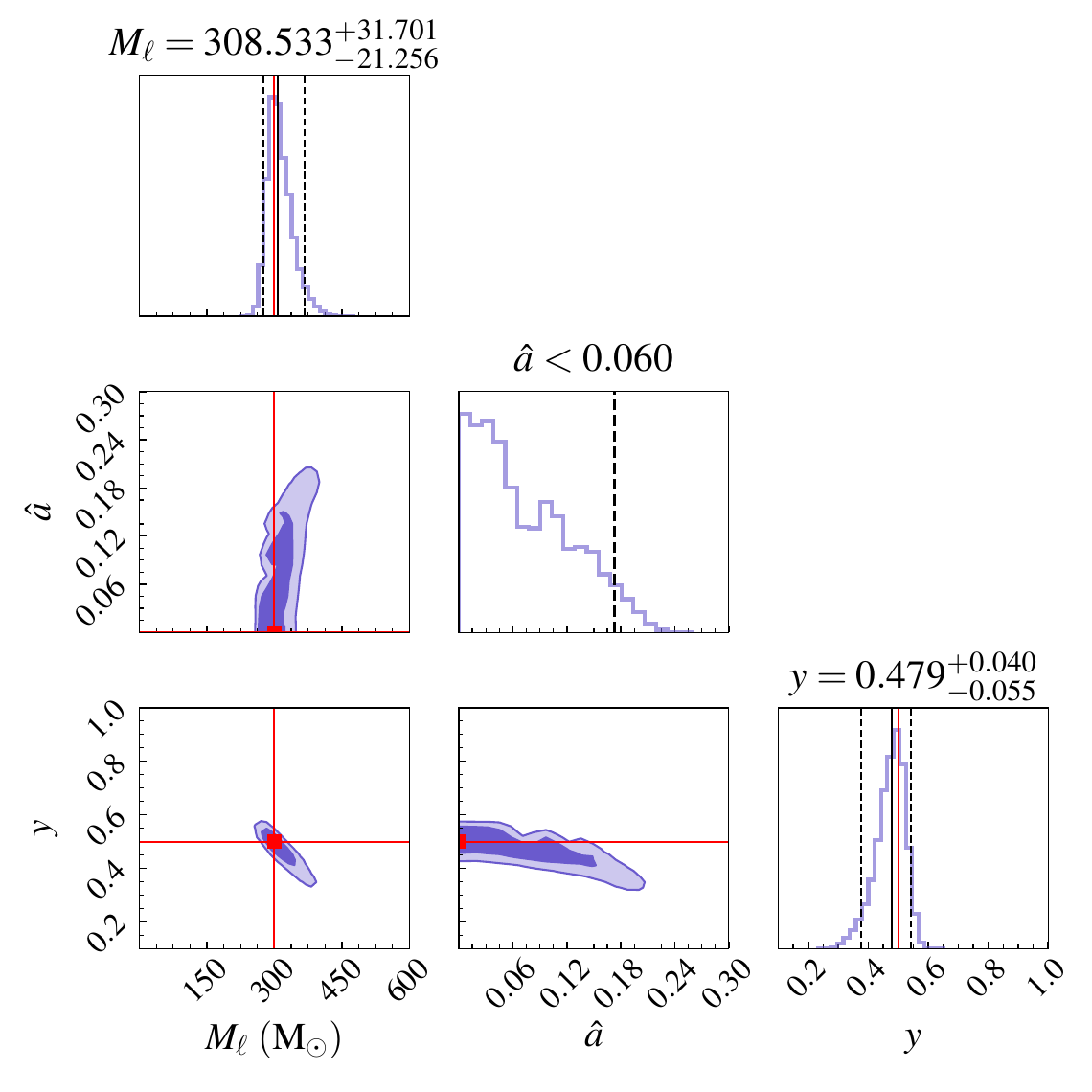}
  \end{tabular}
  \caption{Posterior distributions for the case of rotating (top row) and non-rotating (bottom row) lensed waveforms, with $\text{SNR}=50$ (left column) and $\text{SNR}=100$ (right column), considering ET-D sensitivity. For the rotating case, the true lens parameters (red lines) are [$\hat{a}=0.22,\,M_\ell=300M_{\odot},\, \mathrm{y}=0.5$], with the same $M_\ell$, and $\,y$ values for the zero-spin case. In all cases, the projected spin-angle is fixed at $\phi=0$. The vertical black dashed lines show the 90\% confidence interval, while the vertical black solid lines represent the median value. The unlensed waveform is a quasi-circular signal from a $20-20 M_\odot$ binary at redshift $z_s=2$.}
  \label{fig:recovery_grid}
\end{figure*}

As we saw in the previous sections, in general, the spinning lens can produce up to five images (see, e.g., the top-right panel in Fig.~\ref{fig:waveforms_combined}), but, as a proof of concept, we restrict our analysis to lensing scenarios which can produce up to three images in the geometric optics approximation. We assume flat priors in $\hat{a}\in [0, 0.3], M_\ell\in[100, 600]M_\odot$ and $\mathrm{y}\in[0.1, 1]$~\footnote{The upper limit of $\hat{a}=0.3$ is chosen because we stick to the slowly rotating regime, and beyond $\hat{a}=0.3$, the next higher order spin correction at order $O(\hat{a}^2)$ starts to contribute significantly.}. Hence, the posterior $p(\vec{\theta}_\ell\,|\,d)$ is essentially the same as the approximate Bayesian likelihood $\mathcal{L}(\vec{\theta}_\ell)$ given in Eq.~\eqref{eq:lkhd_high-SNR}. 
Using this, we perform PE for two different true lensed GW waveforms as shown in the corner plots of Fig.~\ref{fig:recovery_grid}. In the top row of Fig.~\ref{fig:recovery_grid}, the true waveform is a GW lensed by a rotating lens with parameters: [$\hat{a}=0.22, \, M_\ell = 300M_{\odot}, \, \mathrm{y}=0.5$] and the second row is for a non-spinning lens with true lens parameters: [$\hat{a}=0, \, M_\ell = 300M_{\odot}, \, \mathrm{y}=0.5$]. In both cases, the unlensed waveforms are assumed to be due to an equal mass, non-spinning binary of component mass $20 M_\odot$ at redshift $z_s=2$. The left and right columns are for SNRs $50$ and $100$, respectively. The one-dimensional marginalised posteriors are also shown in each of the plots, in which we show the 5\textsuperscript{th}, 50\textsuperscript{th} and 95\textsuperscript{th} quantiles as vertical dashed lines. 

From the top row panel of Fig.~\ref{fig:recovery_grid}, we find that the true parameter values of the spinning lens are recovered within the $90\%$ credible interval. Additionally, as expected, an increase in SNR makes the error bars shrink. A few features are to be noted -- e.g, there is a clear correlation between $M_\ell$ and $\mathrm{y}$. This can be understood in terms of the Fresnel integral Eq.~\eqref{eq:ampfac_dimensionless_fd}. Increasing $M_\ell$ leads to an increase of $w$, while increasing $\mathrm{y}$ increases the relative time delays $\tau_\mathrm{d}$ -- both leading to an increase in the magnitude of the exponent in the integral.
On the other hand, there is no strong correlation between $\hat{a}$ with either $M_\ell$ or $\mathrm{y}$. 
Similarly, in the bottom row panel of Fig.~\ref{fig:recovery_grid}, we show the posterior distributions of a lensed GW signal from a non-rotating lens, using rotating lens template waveforms for recovery. Similar to the non-zero spin injection, we note that we can recover the lens parameters within the $90\%$ credible intervals for both SNRs. The correlations can also be seen in this case between $M_\ell$ and $\mathrm{y}$. We show that the upper bounds on the spin can be put at the level of $\hat{a}\lesssim 0.1$ for relatively low SNR, while for larger SNR values, the upper bound can be as good as $\hat{a}\lesssim 0.06$. 

\section{Summary and Outlook}\label{sec:summary}

Most objects in the universe are expected to have non-zero rotation. The rotation of such objects might have been shaped by astrophysically rich processes or possibly originated even from exotic hypothetical objects. Hence, it is of high interest to probe the spins of such objects. In the context of gravitational lensing of GWs in the wave-optics regime, the lensed GW carries information about the structure of the lens, such as, for example, its charge and spin. Hence, future observations of lensed GWs can be used to probe the rotation of isolated compact objects. 

In this work, we analytically calculate the time delay function for lensing of GWs via a slowly rotating black hole in the weak field regime. Using the time delay function, we numerically evaluate the lensed GW waveforms using methods based on~\cite{Deka:2025vzx}. We assess, for the first time, the parameter estimation of GWs lensed by a slowly rotating BH in the weak field regime. We do this for future detector sensitivities; more particularly, we use the ET sensitivity curve in the ET-D configuration~\cite{Hild2011ETsensitivity}. We find that for astrophysical BHs, the rotation corrections to the GW waveforms are extremely tiny, virtually ruling out the possibility of detecting spin using the methods discussed in this work. Hence, we allow for spin values that are not bounded to be extremal, as found in naked singularity solutions both within GR, and in beyond-GR theories. 

We find that the resulting signatures on the GW waveform are non-trivial and strong enough that the lensed GW waveform looks very different from the unlensed GW signal.
For numerical feasibility, we perform parameter estimation using a high-SNR approximation for the likelihood. We estimate the lens parameters using this procedure for two different true lensed waveform injections: GWs from a lens with rotation and without rotation. In both the above cases, we find that the lens parameters are well recovered within the $90\%$ confidence intervals, especially for SNR$=100$. This shows that if such hypothetical compact objects exist in our universe, we can confidently detect their rotation using future observations of gravitationally lensed GWs in the weak field regime.

Our work restricted itself to the slow-rotation, weak-field regime. Moreover, parameter estimation assumed that the source and lens parameters are uncorrelated. In future work, we plan to relax these assumptions to enable parameter estimation for larger spins, while also studying the correlation between source and lens parameters by performing a large-scale sampling of the full GW parameter estimation likelihood. 
\begin{acknowledgments}
We thank Souvik Jana for his careful review and valuable feedback. The research of SC is supported by MATRICS (MTR/2023/000049) and Core Research Grants (CRG/2023/000934) from SERB, ANRF, Government of India. SC also thanks the local hospitality at ICTS and IUCAA through the associateship program, where a part of this work was done.  UD acknowledges the support of the Department of Atomic Energy, Government of India, under project no. RTI4001. SJK gratefully acknowledges support from ANRF/SERB grants SRG/2023/000419 and MTR/2023/000086. This research was supported in part by the International Centre for Theoretical Sciences (ICTS) for participating in the program - The Future of Gravitational-Wave Astronomy 2025 (code: ICTS/FGWA2025/10). 
\end{acknowledgments}
\appendix
\section{Numerical calculation of the magnification function}\label{sec:Numcal_amp_fac}

As mentioned in section \ref{sec:methods}B, the integrand in Eq.~\eqref{eq:ampfac_dimensionless_fd} can become highly oscillatory for high $\mathit{w}\tau_{\rm d}(\vec{x},\vec{\mathrm{y}})$, which makes it numerically challenging to compute the magnification function directly. Hence, following \cite{Deka:2024ecp, Deka:2025vzx}, we calculate the magnification function in the following method:
\begin{enumerate}
    \item Define $\Ft$ as the inverse Fourier transform of $\left[2\pi i \Fw/w\right]$. With a little algebra, we then have:
    \begin{equation}
  \Ft =\int d^2\vec{x}\;\delta\left[\tau_{\rm d}(\vec{x}, \vec{\mathrm{y}})-t\right],\label{eq:ampfac_td}
\end{equation}
This implies that $\Ft$ can be calculated by integrating the time delay on the lens plane over surfaces of constant time delays.

\item The analytic form for $\Ft$
which is valid around the saddle peaks is used to replace the numerically calculated parts of $\Ft$ around it's saddle peaks to improve accuracy, which we call peak reconstruction (see the next section)    
\item  $\Fw$ is then defined by the Fourier transform of $\Ft$ :
\begin{equation}
  \label{eq:FwFFT}
  \Fw = \frac{\mathit{w}}{2\pi i}\times\rm{FFT}\left[\Ft\right]~,
\end{equation}
where $\text{FFT}[\cdot]$ refers to the fast Fourier transform.

\end{enumerate}
The standard procedure used in \cite{Deka:2025vzx} is to sample the time delays on a uniform grid in the lens plane, and calculate the histogram of the samples to calculate $\Ft$. For the case of the rotating lens, however, we can have up to five images. As shown in the right-most figure of the top panel in Fig.~\ref{fig:waveforms_combined}, for the five-image case, it might be that one of the saddle peaks is at an extremely close time relative to the global minima (the saddle peak around $t\approx10^{-6}$ ). In such cases, the sampling of the time delay function in the lens plane should have sufficient sampling density to capture features at such low relative time delays. This can be a computational bottleneck if we use the standard uniform sampling procedure described above. Hence, we adopt a sampling technique which is denser around the region in the lens plane corresponding to the image location of the saddle peak at very low relative time delays. The full $\Ft$  is then obtained by calculating a re-weighted histogram whose weights are proportional to the sampling density in any region. This idea can also be extended to sample any region of the lens plane, which requires more accuracy.

\subsubsection*{Peak reconstruction}\label{sec:appex_A:PR}
The time domain magnification function Eq.~\eqref{eq:ampfac_td} has logarithmically diverging peaks at the location in time of the saddle images. This poses a challenge for the numerical computation of $\Ft$ at and around the time of the saddle images, because it demands higher resolution, especially for higher $w$ waveforms, which is the case for higher lens masses. Hence, to alleviate this computational problem \cite{Deka:2025vzx} use the analytical formula for the logarithmic part of $\Ft$, which is valid around the saddle peaks:
\begin{equation}\label{eq:Ft_near_peak}
    \widetilde{\mathcal{F}}_{\rm approx}(t)=\sqrt{\mu_{+}}-\frac{1}{\pi}\sqrt{|\mu_{-}|}\;\ln\left(|t - t_{\rm peak}|\right) ~,
\end{equation}
where $\mu_{-}$ and $\mu_{+}$ are the GO saddle and minima image magnifications, respectively.
This formula can be extended to the case with more than two lensed images. If we have $n_S$ and $n_M$ saddle and minima images respectively, we can generalize Eq.~\eqref{eq:Ft_near_peak} as:
\begin{equation}\label{eq:Ft_near_peak_general}
    \widetilde{\mathcal{F}}_{\rm approx}(t)=\sum_{i=1}^{n_M}\sqrt{\mu^i_{+}}-\frac{1}{\pi}\sum_{j=1}^{n_S}\sqrt{|\mu^j_{-}|}\;\ln\left(|t - t^j_{\rm peak}|\right) ~,
\end{equation}

Using this analytical formula, the region of the numerically calculated $\Ft$ is replaced by the analytically calculated $\Ft_{\rm approx}$, in some fraction of the region around the saddle peaks, called the peak reconstruction window. Since the width and divergence of the peaks depend on their individual magnifications, for different saddle peaks, we set the window fractions relative to their respective magnifications. \\
For the case of the rotating lens, we can have up to five images (three saddle and two minima). Hence, for more numerical accuracy, we use peak reconstruction to improve the accuracy for all the saddle peaks. 

\section{Expected likelihood in the high-SNR limit}\label{sec:lkhd_high-SNR_justification}
In this Section, we summarise the key steps leading to the noise-averaged likelihood in the high-SNR limit. 

We start by assuming that the detector output is of the form
\begin{equation}
    d(f) = h^{\mathrm{tr}}(f) + n(f)~,
\end{equation}
consisting of a true signal $h^\mathrm{tr}(f)\equiv h(f;\vec{\theta}^\mathrm{tr})$ with signal parameters $\vec{\theta}^\mathrm{tr}$, and a specific realisation of a stationary, zero-mean Gaussian noise $n(f)$. The PSD is related to the noise as~\cite{Cutler:1994ys}:
\begin{equation}
    \langle n^\ast(f)\,n(f^\prime)\rangle_n = \frac{1}{2} S_n(f)\,\delta(f-f^\prime)~,
\end{equation}
where $\langle\cdots\rangle_n$ denotes the ensemble average over many possible noise realisations.

Under the hypothesis $\mathcal{H}$ that the data contains a signal $h(\vec{\theta})\equiv h(f;\vec{\theta}, \mathcal{H})$, the likelihood can be defined in terms of the inner product (see Eq.~\eqref{eq:inner_product}) as~\cite{Cutler:1994ys}
\begin{equation}\label{eq:lkhd_gaussian}
    p(d\,|\,\vec{\theta}, \mathcal{H})\propto\exp\left[-\frac{1}{2}\langle d-h(\vec{\theta}), d-h(\vec{\theta})\rangle\right]~.
\end{equation}

Evaluating the expectation value of the log-likelihood over noise realisations, and using the properties of Gaussian noise, we obtain
\begin{subequations}\label{eq:lkhd_avg}
\begin{align}
    \langle\log p(d\,|\,\vec{\theta}, \mathcal{H})\rangle_n=&-\frac{1}{2}\langle\langle d-h(\vec{\theta}), d-h(\vec{\theta})\rangle\rangle_n + \mathrm{const.}\\
    =&-\frac{1}{2}\left[\langle h^\mathrm{tr}-h(\vec{\theta}), h^\mathrm{tr}-h(\vec{\theta})\rangle\right]\nonumber\\ 
    & - \frac{1}{2}\langle\langle n,n\rangle\rangle_n+ \mathrm{const.}~
\end{align}
\end{subequations}
The noise term $\langle\langle n,n\rangle\rangle_n$ is independent of $\vec{\theta}$ and can be absorbed into the normalisation.

The optimal SNR for the true signal is defined as 
\begin{equation}\label{eq:opt_SNR}
    \rho^2\equiv\langle h^\mathrm{tr}, h^\mathrm{tr}\rangle~.
\end{equation}
If the template is normalised such that $\langle h(\vec{\theta}), h(\vec{\theta})\rangle\simeq \langle h^\mathrm{tr}, h^\mathrm{tr}\rangle =\rho^2$, then
\begin{equation}
    \langle h^\mathrm{tr}, h(\vec{\theta})\rangle = \rho^2\left(1-\mathcal{M}(\vec{\theta}^\mathrm{tr}, \vec{\theta})\right)~,
\end{equation}
where $\mathcal{M}$ is the mismatch (see Eq.~\eqref{eq:match_definition}). Expanding the inner product in Eq.~\eqref{eq:lkhd_avg} then gives
\begin{subequations}
\begin{align}
    \langle h^\mathrm{tr}-h(\vec{\theta}), h^\mathrm{tr}-h(\vec{\theta})\rangle = & \langle h^\mathrm{tr}, h^\mathrm{tr}\rangle +\langle h(\vec{\theta}), h(\vec{\theta})\rangle\nonumber\\
    & - 2\langle h^\mathrm{tr}, h(\vec{\theta})\rangle\\
    = & 2\rho^2\mathcal{M}(\vec{\theta}^\mathrm{tr}, \vec{\theta})~.
\end{align}
\end{subequations}
Therefore, the expected likelihood becomes
\begin{equation}
    \langle p(d\,|\,\vec{\theta}, \mathcal{H})\rangle_n\propto\exp\left[-\rho^2\mathcal{M}(\vec{\theta}^\mathrm{tr}, \vec{\theta})\right]
\end{equation}

\bibliography{References}

@article{Baraldo:1999ny,
    author = "Baraldo, Christian and Hosoya, Akio and Nakamura, Takahiro T.",
    title = "{Gravitationally induced interference of gravitational waves by a rotating massive object}",
    doi = "10.1103/PhysRevD.59.083001",
    journal = "Phys. Rev. D",
    volume = "59",
    pages = "083001",
    year = "1999"
}

@article{Asada:2000vn,
    author = "Asada, Hideki and Kasai, Masumi",
    title = "{Can we see a rotating gravitational lens?}",
    eprint = "astro-ph/0006157",
    archivePrefix = "arXiv",
    doi = "10.1143/PTP.104.95",
    journal = "Prog. Theor. Phys.",
    volume = "104",
    pages = "95",
    year = "2000"
}

@book{Misner:1973prb,
    author = "Misner, Charles W. and Thorne, K. S. and Wheeler, J. A.",
    title = "{Gravitation}",
    isbn = "978-0-7167-0344-0, 978-0-691-17779-3",
    publisher = "W. H. Freeman",
    address = "San Francisco",
    year = "1973"
}

@book{Schneider:1992bmb,
    author = {Schneider, Peter and Ehlers, J{\"u}rgen and Falco, Emilio E.},
    title = "{Gravitational Lenses}",
    doi = "10.1007/978-3-662-03758-4",
    isbn = "978-3-540-66506-9, 978-3-662-03758-4",
    publisher = "Springer",
    series = "Astronomy and Astrophysics Library",
    year = "1992"
}

@article{McClintock:2013vwa,
    author = "McClintock, Jeffrey E. and Narayan, Ramesh and Steiner, James F.",
    title = "{Black Hole Spin via Continuum Fitting and the Role of Spin in Powering Transient Jets}",
    eprint = "1303.1583",
    archivePrefix = "arXiv",
    primaryClass = "astro-ph.HE",
    doi = "10.1007/s11214-013-0003-9",
    journal = "Space Sci. Rev.",
    volume = "183",
    pages = "295--322",
    year = "2014"
}

@article{Fabian2000,
  author  = {Fabian, A. C. and Iwasawa, K. and Reynolds, C. S. and Young, A. J.},
  title   = {Broad iron lines in active galactic nuclei},
  journal = {Publications of the Astronomical Society of the Pacific},
  year    = {2000},
  volume  = {112},
  pages   = {1145--1161},
  doi     = {10.1086/316605},
  eprint  = {arXiv:astro-ph/0004366}
}

@misc{vanderKlis1997,
  author  = {van der Klis, M.},
  title   = {Kilohertz Quasi-Periodic Oscillations in Low-Mass X-Ray Binaries:
             A Review},
  howpublished = {arXiv:astro-ph/9710016},
  year    = {1997},
  note    = {Review (see also later 2006 summaries)}
}

@article{Akiyama2019,
  author  = {Akiyama, Kazunori and Event Horizon Telescope Collaboration and others},
  title   = {First M87 Event Horizon Telescope Results. I. The Shadow of the
             Supermassive Black Hole},
  journal = {Astrophysical Journal Letters},
  year    = {2019},
  volume  = {875},
  pages   = {L1},
  doi     = {10.3847/2041-8213/ab0ec7}
}

@article{Eigenbrod2008,
  author  = {Eigenbrod, A. and Courbin, F. and Sluse, D. and Meylan, G. and Agol, E.},
  title   = {Microlensing variability in the gravitationally lensed quasar
             Q2237+0305},
  journal = {Astronomy \& Astrophysics},
  year    = {2008},
  volume  = {490},
  pages   = {933--943},
  doi     = {10.1051/0004-6361:200809686},
  eprint  = {arXiv:0709.2828}
}

@article{Pooley2007,
  author  = {Pooley, D. and Blackburne, J. A. and Rappaport, S. and Schechter, P. L.},
  title   = {X-Ray and Optical Flux Ratio Anomalies in Quadruply Lensed Quasars},
  journal = {Astrophysical Journal},
  year    = {2007},
  volume  = {661},
  pages   = {19--29},
  doi     = {10.1086/513524}
}

@book{Schneider1992,
  author    = {Schneider, P. and Ehlers, J. and Falco, E. E.},
  title     = {Gravitational Lenses},
  series    = {Astronomy and Astrophysics Library},
  publisher = {Springer},
  year      = {1992},
  doi       = {10.1007/978-1-4612-2756-4}
}

@article{Nakamura1998,
  author  = {Nakamura, T. T.},
  title   = {Gravitational lensing of gravitational waves from inspiraling binaries},
  journal = {Physical Review Letters},
  year    = {1998},
  volume  = {80},
  pages   = {1138--1141},
  doi     = {10.1103/PhysRevLett.80.1138}
}

@article{Kopeikin2002,
  author  = {Kopeikin, S. and Mashhoon, B.},
  title   = {Gravitomagnetic effects in the propagation of electromagnetic waves in variable gravitational fields of arbitrary-moving and spinning bodies},
  journal = {Physical Review D},
  year    = {2002},
  volume  = {65},
  pages   = {064025},
  doi     = {10.1103/PhysRevD.65.064025},
  eprint  = {arXiv:gr-qc/0110101}
}

@article{Sereno2003,
  author  = {Sereno, M.},
  title   = {Gravitational lensing in metric theories of gravity},
  journal = {Physical Review D},
  year    = {2003},
  volume  = {67},
  pages   = {064007},
  doi     = {10.1103/PhysRevD.67.064007},
  eprint  = {arXiv:astro-ph/0301290}
}

@article{Sereno2005,
  author  = {Sereno, M.},
  title   = {On gravitomagnetic time-delay by extended lenses},
  journal = {Monthly Notices of the Royal Astronomical Society},
  year    = {2005},
  volume  = {357},
  number  = {4},
  pages   = {1205--1210},
  doi     = {10.1111/j.1365-2966.2005.08709.x},
  eprint  = {arXiv:astro-ph/0412108}
}

@article{Takahashi2003,
  author  = {Takahashi, R. and Nakamura, T.},
  title   = {Wave effects in the gravitational lensing of gravitational waves},
  journal = {Astrophysical Journal},
  year    = {2003},
  volume  = {595},
  pages   = {1039--1051},
  doi     = {10.1086/377430},
  eprint  = {arXiv:astro-ph/0305055}
}

@article{Heger2005,
  author  = {Heger, A. and Woosley, S. E. and Spruit, H. C.},
  title   = {Presupernova evolution of differentially rotating massive stars including magnetic fields},
  journal = {Astrophysical Journal},
  year    = {2005},
  volume  = {626},
  pages   = {350--363},
  doi     = {10.1086/429868},
  eprint  = {astro-ph/0409422}
}

@article{Fuller2019,
  author  = {Fuller, J. and Ma, L.},
  title   = {Most black holes are born very slowly rotating},
  journal = {Astrophysical Journal Letters},
  year    = {2019},
  volume  = {881},
  pages   = {L1},
  doi     = {10.3847/2041-8213/ab2a4f},
  eprint  = {arXiv:1907.03714}
}

@article{Bonga:2024orc,
    author = "Bonga, B{\'e}atrice and Feldbrugge, Job and Ribes Metidieri, Ariadna",
    title = "{Wave optics for rotating stars}",
    eprint = "2410.03828",
    archivePrefix = "arXiv",
    primaryClass = "gr-qc",
    doi = "10.1103/PhysRevD.111.063061",
    journal = "Phys. Rev. D",
    volume = "111",
    number = "6",
    pages = "063061",
    year = "2025"
}

@article{Deka:2024ecp,
    author = "Deka, Uddeepta and Chakraborty, Sumanta and Kapadia, Shasvath J. and Shaikh, Md Arif and Ajith, Parameswaran",
    title = "{Probing the charge of compact objects with gravitational microlensing of gravitational waves}",
    eprint = "2401.06553",
    archivePrefix = "arXiv",
    primaryClass = "gr-qc",
    doi = "10.1103/PhysRevD.111.064028",
    journal = "Phys. Rev. D",
    volume = "111",
    number = "6",
    pages = "064028",
    year = "2025"
}

@article{Deka:2025vzx,
    author = "Deka, Uddeepta and Prabhu, Gopalkrishna and Shaikh, Md Arif and Kapadia, Shasvath J. and Varma, Vijay and Field, Scott E.",
    title = "{Surrogate modeling of gravitational waves microlensed by spherically symmetric potentials}",
    eprint = "2501.02974",
    archivePrefix = "arXiv",
    primaryClass = "gr-qc",
    doi = "10.1103/PhysRevD.111.104042",
    journal = "Phys. Rev. D",
    volume = "111",
    number = "10",
    pages = "104042",
    year = "2025"
}

@misc{iyer2009strongweakdeflectionlight,
      title={Strong and Weak Deflection of Light in the Equatorial Plane of a Kerr Black Hole}, 
      author={Savitri V. Iyer and Edward C. Hansen},
      year={2009},
      eprint={0908.0085},
      archivePrefix={arXiv},
      primaryClass={gr-qc},
      url={https://arxiv.org/abs/0908.0085}, 
}

@article{Reynolds:2013qqa,
    author = "Reynolds, Christopher S.",
    title = "{Measuring Black Hole Spin using X-ray Reflection Spectroscopy}",
    eprint = "1302.3260",
    archivePrefix = "arXiv",
    primaryClass = "astro-ph.HE",
    doi = "10.1007/s11214-013-0006-6",
    journal = "Space Sci. Rev.",
    volume = "183",
    number = "1-4",
    pages = "277--294",
    year = "2014"
}

@article{waveoptics_numerical_methods,
  title = {Lensing of gravitational waves: Efficient wave-optics methods and validation with symmetric lenses},
  author = {Tambalo, Giovanni and Zumalac\'arregui, Miguel and Dai, Liang and Cheung, Mark Ho-Yeuk},
  journal = {Phys. Rev. D},
  volume = {108},
  issue = {4},
  pages = {043527},
  numpages = {20},
  year = {2023},
  month = {Aug},
  publisher = {American Physical Society},
  doi = {10.1103/PhysRevD.108.043527},
  url = {https://link.aps.org/doi/10.1103/PhysRevD.108.043527}
}

@article{Mishra:2023ddt,
    author = "Mishra, Anuj and Meena, Ashish Kumar and More, Anupreeta and Bose, Sukanta",
    title = "{Exploring the impact of microlensing on gravitational wave signals: Biases, population characteristics, and prospects for detection}",
    eprint = "2306.11479",
    archivePrefix = "arXiv",
    primaryClass = "astro-ph.CO",
    doi = "10.1093/mnras/stae836",
    journal = "Mon. Not. Roy. Astron. Soc.",
    volume = "531",
    number = "1",
    pages = "764--787",
    year = "2024"
}

@article{Hild2011ETsensitivity,
  author       = {Hild, S. and Abernathy, M. and Acernese, F. and Amaro-Seoane, P. and Andersson, N. and Arun, K. and Barone, F. and Barr, B. and Barsuglia, M. and Beker, M.},
  title        = {Sensitivity studies for third-generation gravitational wave observatories},
  journal      = {Classical and Quantum Gravity},
  volume       = {28},
  number       = {9},
  pages        = {094013},
  year         = {2011},
  doi          = {10.1088/0264-9381/28/9/094013},
  url          = {https://doi.org/10.1088/0264-9381/28/9/094013},
  publisher    = {IOP Publishing}
}

@article{aligo,
    author = "Aasi, J. and others",
    collaboration = "LIGO Scientific",
    title = "{Advanced LIGO}",
    eprint = "1411.4547",
    archivePrefix = "arXiv",
    primaryClass = "gr-qc",
    doi = "10.1088/0264-9381/32/7/074001",
    journal = "Class. Quant. Grav.",
    volume = "32",
    pages = "074001",
    year = "2015"
}

@article{avirgo,
    author = "Acernese, F. and others",
    collaboration = "VIRGO",
    title = "{Advanced Virgo: a second-generation interferometric gravitational wave detector}",
    eprint = "1408.3978",
    archivePrefix = "arXiv",
    primaryClass = "gr-qc",
    doi = "10.1088/0264-9381/32/2/024001",
    journal = "Class. Quant. Grav.",
    volume = "32",
    number = "2",
    pages = "024001",
    year = "2015"
}

@article{KAGRA,
    author = "Akutsu, T. and others",
    collaboration = "KAGRA",
    title = "{Overview of KAGRA: Detector design and construction history}",
    eprint = "2005.05574",
    archivePrefix = "arXiv",
    primaryClass = "physics.ins-det",
    doi = "10.1093/ptep/ptaa125",
    journal = "PTEP",
    volume = "2021",
    number = "5",
    pages = "05A101",
    year = "2021"
}

@article{LIGOScientific:2025slb,
    author = "Abac, A. G. and others",
    collaboration = "LIGO Scientific, VIRGO, KAGRA",
    title = "{GWTC-4.0: Updating the Gravitational-Wave Transient Catalog with Observations from the First Part of the Fourth LIGO-Virgo-KAGRA Observing Run}",
    eprint = "2508.18082",
    archivePrefix = "arXiv",
    primaryClass = "gr-qc",
    reportNumber = "LIGO-P2400386",
    month = "8",
    year = "2025"
}

@article{LIGOScientific:2023bwz,
    author = "Abbott, R. and others",
    collaboration = "LIGO Scientific, KAGRA, VIRGO",
    title = "{Search for Gravitational-lensing Signatures in the Full Third Observing Run of the LIGO{\textendash}Virgo Network}",
    eprint = "2304.08393",
    archivePrefix = "arXiv",
    primaryClass = "gr-qc",
    reportNumber = "LIGO-P2200031",
    doi = "10.3847/1538-4357/ad3e83",
    journal = "Astrophys. J.",
    volume = "970",
    number = "2",
    pages = "191",
    year = "2024"
}

@article{Magare:2023hgs,
    author = "Magare, Sourabh and Kapadia, Shasvath J. and More, Anupreeta and Singh, Mukesh Kumar and Ajith, Parameswaran and Ramprakash, A. N.",
    title = "{Gear Up for the Action Replay: Leveraging Lensing for Enhanced Gravitational-wave Early Warning}",
    eprint = "2302.02916",
    archivePrefix = "arXiv",
    primaryClass = "astro-ph.HE",
    doi = "10.3847/2041-8213/acf668",
    journal = "Astrophys. J. Lett.",
    volume = "955",
    number = "2",
    pages = "L31",
    year = "2023"
}

@article{Magare:2025ulm,
    author = "Magare, Sourabh and More, Anupreeta and Kapadia, Shasvath J.",
    title = "{Early warning for lensed gravitational wave counterparts from time delays of their host galaxies observed in the optical}",
    eprint = "2509.07967",
    archivePrefix = "arXiv",
    primaryClass = "astro-ph.HE",
    month = "9",
    year = "2025"
}

@article{Jana:2024dhc,
    author = "Jana, Souvik and Kapadia, Shasvath J. and Venumadhav, Tejaswi and More, Surhud and Ajith, Parameswaran",
    title = "{Probing the Nature of Dark Matter Using Strongly Lensed Gravitational Waves from Binary Black Holes}",
    eprint = "2408.05290",
    archivePrefix = "arXiv",
    primaryClass = "astro-ph.CO",
    doi = "10.1103/7q31-3qwz",
    journal = "Phys. Rev. Lett.",
    volume = "135",
    number = "11",
    pages = "111402",
    year = "2025"
}

@article{Gerosa2021,
  author       = {Gerosa, Davide and Fishbach, Maya},
  title        = {Hierarchical mergers of stellar-mass black holes and their gravitational-wave signatures},
  journal      = {Nature Astronomy},
  year         = {2021},
  volume       = {5},
  number       = {8},
  pages        = {749--760},
  doi          = {10.1038/s41550-021-01398-w},
  url          = {https://doi.org/10.1038/s41550-021-01398-w},
  issn         = {2397-3366}
}

@article{Jana:2024uta,
    author = "Jana, Souvik and Kapadia, Shasvath J. and Venumadhav, Tejaswi and More, Surhud and Ajith, Parameswaran",
    title = "{Strong-lensing cosmography using third-generation gravitational-wave detectors}",
    eprint = "2405.17805",
    archivePrefix = "arXiv",
    primaryClass = "gr-qc",
    doi = "10.1088/1361-6382/ad8d2e",
    journal = "Class. Quant. Grav.",
    volume = "41",
    number = "24",
    pages = "245010",
    year = "2024"
}

@article{Jana:2022shb,
    author = "Jana, Souvik and Kapadia, Shasvath J. and Venumadhav, Tejaswi and Ajith, Parameswaran",
    title = "{Cosmography Using Strongly Lensed Gravitational Waves from Binary Black Holes}",
    eprint = "2211.12212",
    archivePrefix = "arXiv",
    primaryClass = "astro-ph.CO",
    reportNumber = "LIGO-P2200298",
    doi = "10.1103/PhysRevLett.130.261401",
    journal = "Phys. Rev. Lett.",
    volume = "130",
    number = "26",
    pages = "261401",
    year = "2023"
}

@article{Goyal:2020bkm,
    author = "Goyal, Srashti and Haris, K. and Mehta, Ajit Kumar and Ajith, Parameswaran",
    title = "{Testing the nature of gravitational-wave polarizations using strongly lensed signals}",
    eprint = "2008.07060",
    archivePrefix = "arXiv",
    primaryClass = "gr-qc",
    reportNumber = "LIGO-P2000295-v1",
    doi = "10.1103/PhysRevD.103.024038",
    journal = "Phys. Rev. D",
    volume = "103",
    number = "2",
    pages = "024038",
    year = "2021"
}

@article{Collett:2016dey,
    author = "Collett, Thomas E. and Bacon, David",
    title = "{Testing the speed of gravitational waves over cosmological distances with strong gravitational lensing}",
    eprint = "1602.05882",
    archivePrefix = "arXiv",
    primaryClass = "astro-ph.HE",
    doi = "10.1103/PhysRevLett.118.091101",
    journal = "Phys. Rev. Lett.",
    volume = "118",
    number = "9",
    pages = "091101",
    year = "2017"
}

@article{Basak:2022fig,
    author = "Basak, Soummyadip and Sharma, Aditya Kumar and Kapadia, Shasvath J. and Ajith, Parameswaran",
    title = "{Prospects for the Observation of Continuous Gravitational Waves from Spinning Neutron Stars Lensed by the Galactic Supermassive Black Hole}",
    eprint = "2205.00022",
    archivePrefix = "arXiv",
    primaryClass = "gr-qc",
    doi = "10.3847/2041-8213/acab50",
    journal = "Astrophys. J. Lett.",
    volume = "942",
    number = "2",
    pages = "L31",
    year = "2023"
}

@article{Basak:2021ten,
    author = "Basak, S. and Ganguly, A. and Haris, K. and Kapadia, S. and Mehta, A. K. and Ajith, P.",
    title = "{Constraints on Compact Dark Matter from Gravitational Wave Microlensing}",
    eprint = "2109.06456",
    archivePrefix = "arXiv",
    primaryClass = "gr-qc",
    reportNumber = "LIGO-P2100321",
    doi = "10.3847/2041-8213/ac4dfa",
    journal = "Astrophys. J.",
    volume = "926",
    number = "2",
    pages = "L28",
    year = "2022"
}

@article{Mishra:2021xzz,
    author = "Mishra, Anuj and Meena, Ashish Kumar and More, Anupreeta and Bose, Sukanta and Bagla, Jasjeet Singh",
    title = "{Gravitational lensing of gravitational waves: effect of microlens population in lensing galaxies}",
    eprint = "2102.03946",
    archivePrefix = "arXiv",
    primaryClass = "astro-ph.CO",
    doi = "10.1093/mnras/stab2875",
    journal = "Mon. Not. Roy. Astron. Soc.",
    volume = "508",
    number = "4",
    pages = "4869--4886",
    year = "2021"
}

@article{Punturo:2010zz,
    author = "Punturo, M. and others",
    editor = "Ricci, Fulvio",
    title = "{The Einstein Telescope: A third-generation gravitational wave observatory}",
    doi = "10.1088/0264-9381/27/19/194002",
    journal = "Class. Quant. Grav.",
    volume = "27",
    pages = "194002",
    year = "2010"
}

@article{Barsode:2024wda,
    author = "Barsode, A. and Kapadia, S. J. and Ajith, P.",
    title = "{Constraints on Compact Dark Matter from the Nonobservation of Gravitational-wave Strong Lensing}",
    eprint = "2405.15878",
    archivePrefix = "arXiv",
    primaryClass = "gr-qc",
    doi = "10.3847/1538-4357/ad77c4",
    journal = "Astrophys. J.",
    volume = "975",
    number = "1",
    pages = "48",
    year = "2024"
}

@article{Diego:2019lcd,
    author = "Diego, J. M. and Hannuksela, O. A. and Kelly, P. L. and Broadhurst, T. and Kim, K. and Li, T. G. F. and Smoot, G. F. and Pagano, G.",
    title = "{Observational signatures of microlensing in gravitational waves at LIGO/Virgo frequencies}",
    eprint = "1903.04513",
    archivePrefix = "arXiv",
    primaryClass = "astro-ph.CO",
    doi = "10.1051/0004-6361/201935490",
    journal = "Astron. Astrophys.",
    volume = "627",
    pages = "A130",
    year = "2019"
}

@article{Meena:2019ate,
    author = "Meena, Ashish Kumar and Bagla, J. S.",
    title = "{Gravitational lensing of gravitational waves: wave nature and prospects for detection}",
    eprint = "1903.11809",
    archivePrefix = "arXiv",
    primaryClass = "astro-ph.CO",
    doi = "10.1093/mnras/stz3509",
    journal = "Mon. Not. Roy. Astron. Soc.",
    volume = "492",
    number = "1",
    pages = "1127--1134",
    year = "2020"
}

@article{Cutler:1994ys,
    author = "Cutler, Curt and Flanagan, Eanna E.",
    title = "{Gravitational waves from merging compact binaries: How accurately can one extract the binary's parameters from the inspiral wave form?}",
    eprint = "gr-qc/9402014",
    archivePrefix = "arXiv",
    reportNumber = "GRP-369",
    doi = "10.1103/PhysRevD.49.2658",
    journal = "Phys. Rev. D",
    volume = "49",
    pages = "2658--2697",
    year = "1994"
}

@article{GWTC1,
  title = {GWTC-1: A Gravitational-Wave Transient Catalog of Compact Binary Mergers Observed by LIGO and Virgo during the First and Second Observing Runs},
  author = {Abbott, B. P. and others},
  collaboration = {LIGO Scientific Collaboration and Virgo Collaboration},
  journal = {Phys. Rev. X},
  volume = {9},
  issue = {3},
  pages = {031040},
  numpages = {49},
  year = {2019},
  month = {Sep},
  publisher = {American Physical Society},
  doi = {10.1103/PhysRevX.9.031040},
  url = {https://link.aps.org/doi/10.1103/PhysRevX.9.031040}
}

@article{GWTC2,
  title = {GWTC-2: Compact Binary Coalescences Observed by LIGO and Virgo during the First Half of the Third Observing Run},
  author = {Abbott, R. and others},
  collaboration = {LIGO Scientific Collaboration and Virgo Collaboration},
  journal = {Phys. Rev. X},
  volume = {11},
  issue = {2},
  pages = {021053},
  numpages = {52},
  year = {2021},
  month = {Jun},
  publisher = {American Physical Society},
  doi = {10.1103/PhysRevX.11.021053},
  url = {https://link.aps.org/doi/10.1103/PhysRevX.11.021053}
}

@article{GWTC2.1,
  title = {GWTC-2.1: Deep extended catalog of compact binary coalescences observed by LIGO and Virgo during the first half of the third observing run},
  author = {Abbott, R. and others},
  collaboration = {The LIGO Scientific Collaboration and the Virgo Collaboration},
  journal = {Phys. Rev. D},
  volume = {109},
  issue = {2},
  pages = {022001},
  numpages = {45},
  year = {2024},
  month = {Jan},
  publisher = {American Physical Society},
  doi = {10.1103/PhysRevD.109.022001},
  url = {https://link.aps.org/doi/10.1103/PhysRevD.109.022001}
}

@article{GWTC3,
  title = {GWTC-3: Compact Binary Coalescences Observed by LIGO and Virgo during the Second Part of the Third Observing Run},
  author = {Abbott, R. and others},
  collaboration = {LIGO Scientific Collaboration, Virgo Collaboration, and KAGRA Collaboration},
  journal = {Phys. Rev. X},
  volume = {13},
  issue = {4},
  pages = {041039},
  numpages = {89},
  year = {2023},
  month = {Dec},
  publisher = {American Physical Society},
  doi = {10.1103/PhysRevX.13.041039},
  url = {https://link.aps.org/doi/10.1103/PhysRevX.13.041039}
}

@article{Banerjee:2021aln,
    author = "Banerjee, Indrani and Chakraborty, Sumanta and SenGupta, Soumitra",
    title = "{Looking for extra dimensions in the observed quasi-periodic oscillations of black holes}",
    eprint = "2105.06636",
    archivePrefix = "arXiv",
    primaryClass = "gr-qc",
    doi = "10.1088/1475-7516/2021/09/037",
    journal = "JCAP",
    volume = "09",
    pages = "037",
    year = "2021"
}

@article{Banerjee:2019nnj,
    author = "Banerjee, Indrani and Chakraborty, Sumanta and SenGupta, Soumitra",
    title = "{Silhouette of M87*: A New Window to Peek into the World of Hidden Dimensions}",
    eprint = "1909.09385",
    archivePrefix = "arXiv",
    primaryClass = "gr-qc",
    doi = "10.1103/PhysRevD.101.041301",
    journal = "Phys. Rev. D",
    volume = "101",
    number = "4",
    pages = "041301",
    year = "2020"
}

@article{Banerjee:2019sae,
    author = "Banerjee, Indrani and Chakraborty, Sumanta and SenGupta, Soumitra",
    title = "{Decoding signatures of extra dimensions and estimating spin of quasars from the continuum spectrum}",
    eprint = "1905.08043",
    archivePrefix = "arXiv",
    primaryClass = "gr-qc",
    doi = "10.1103/PhysRevD.100.044045",
    journal = "Phys. Rev. D",
    volume = "100",
    number = "4",
    pages = "044045",
    year = "2019"
}

@article{Mishra:2019trb,
    author = "Mishra, Akash K. and Chakraborty, Sumanta and Sarkar, Sudipta",
    title = "{Understanding photon sphere and black hole shadow in dynamically evolving spacetimes}",
    eprint = "1903.06376",
    archivePrefix = "arXiv",
    primaryClass = "gr-qc",
    doi = "10.1103/PhysRevD.99.104080",
    journal = "Phys. Rev. D",
    volume = "99",
    number = "10",
    pages = "104080",
    year = "2019"
}

\end{document}